\documentstyle[12pt, aaspp]{article}
\tighten
\singlespace

\begin{document}

\title{Decaying Neutrinos in Galaxy Clusters}
\author{Adrian L. Melott and Randall J. Splinter}
\affil{Department of Physics and Astronomy}
\authoraddr{University of Kansas \\ Lawrence, KS 66045}
\and
\author{Massimo Persic$^*$ and Paolo Salucci}
\affil{SISSA and $^*$ Osservatorio Astronomic}
\authoraddr{Strada Costiera 11, I -- 34014 \\ Trieste, Italy}

\received{June 23, 1993}
\accepted{July 7, 1993}

\begin{abstract}

Davidsen et al. (1991) have argued that
the failure to detect $uv$ photons from the dark matter DM) in cluster
A665 excludes the decaying neutrino  hypothesis.  Sciama et al. (1993)
argued that  because  of high  central concentration  the  DM  in that
cluster  must  be baryonic.  We  study the  DM profile  in clusters of
galaxies simulated using the Harrison--Zel'dovich  spectrum of density
fluctuations,   and an  amplitude  previously  derived  from numerical
simulations (Melott 1984b; Anninos et  al. 1991) and in agreement with
microwave background  fluctuations (Smoot et  al. 1992).  We find that
with this  amplitude normalization  cluster neutrino DM densities  are
comparable to observed cluster DM values.  We conclude that given this
normalization, the cluster DM should  be at least largely composed  of
neutrinos. The constraint of Davidsen et al. can be  somewhat weakened
by the presence of baryonic DM; but it cannot  be eliminated given our
assumptions.

\end{abstract}

%\twocolumn
\section{Introduction}

Over a period spanning  roughly the last ten  years a number of papers
have  examined the astrophysical  and  cosmological consequences of  a
massive neutrino species possessing a specific radiative decay mode to
photons with energies of roughly 15 eV and  at a rate of approximately
$10^{-24}  {\rm s^{-1}}$.    Melott  (1984a)  was  able  to provide an
explanation for the  morphological segregation of  dwarf galaxies near
large parent galaxies as seen by Einasto et al. (1974) by arguing that
near  the  large  parent galaxy  the flux  of   {\it  uv} photons from
neutrino decay would be larger and therefore one would expect that the
satellite  galaxies   found nearer  the  parent galaxy  should exhibit
little  neutral gas  and  hence little   star formation.  While  those
satellite galaxies further away  from  the  parent galaxy  would see a
much smaller {\it uv} flux and hence contain larger amounts of neutral
gas.   In  addition Melott was  able to derive  the slope dividing the
morphological  types fitting,   the observations of    Einasto et  al.
(1974).   In  1988  Melott  et al.   extended   the model  to  provide
explanations for another series of effects. Using a simple equilibrium
argument they were able to show that with  a neutrino of mass $\approx
30$  eV  and lifetime  $\approx  10^{24}$ seconds, that   ionizing the
universe to  those  levels  required   by the  Gunn-Peterson test   is
completely reasonable  in  the  neutrino decay   scenario    (see also
Rephaeli \& Szalay 1981 and Sciama 1982). More  recently Sciama (1990)
has provided several additional problems that the neutrino decay model
can   provide  a  simple  explanation for if  $\tau  \sim (1-3) \times
10^{23}$ sec.

Several attempts have been made for  an  unambiguous detection  of the
{\it uv} flux from the neutrino decay, ever since the earlier theories
of Cowsik  (1977)  and de  Rujula  \& Glashow  (1980) (see  Shipman \&
Cowsik 1981; Henry \& Feldman  1981;  Holberg \& Barber  1985; Fabian,
Naylor  \& Sciama 1991). In  1991 Davidsen et al.,   using the Hopkins
Ultraviolet Telescope,  performed a series of  observations  hoping to
provide  evidence for or   against  Sciama's (1990)  decaying neutrino
model. The  Davidsen  et al. experiment   centered  around  the galaxy
cluster A665.   This cluster is known  to be among  the  richest known
clusters, therefore  one would expect  that it  might  possess a large
dark matter (DM) halo.  The DM halo consisting primarily  of neutrinos
would generate a {\it uv} flux due to the decay of the neutrinos.  The
Davidsen experiment was unable to find any convincing evidence for the
predicted {\it uv} flux at the level anticipated from  the decaying DM
model, and they  concluded that $log  \ \tau  ({\rm seconds}) >  24.5$
roughly. Furthermore they argued that the theory can only remain valid
if one  of the following conditions  holds:(1) the cluster  is several
times  less massive  than estimated in their work  (or  there exists a
significant baryonic dark matter  component) and the  redshifted decay
photon energy happens to lie near the Ly $\beta$  airglow line, or (2)
there is substantial absorption along the line of sight.

Sciama et  al.  (1993) argue that based  upon  several  new  pieces of
evidence  that assuming that all  of  the DM  in the cluster is in the
form  of  neutrinos may perhaps be  an overly restrictive requirement.
The first new piece of  evidence concerns recent X-ray observations of
A665. In 1992   Hughes \&  Tanaka  found strong  evidence that the  DM
distribution in A665 is more   centrally condensed  than  either   the
galaxy distribution or the hot  X-ray emitting  plasma.  Sciama et al.
(1993) argue for  a  baryonic form  of DM  in the  central core of the
cluster because    the  neutrinos  being   nearly  collisionless would
presumably  be incapable of  dissipating their  energy enough to  fall
deeply into the potential of  the cluster.  Secondly, a calculation by
Persic \& Salucci (1992)  of   the contribution  of visual matter   to
$\Omega = \rho/
\rho_{\rm{crit}}$  shows  that  $\Omega_{\rm{vis}} \approx 0.003$.
Comparing this   to     the   value   of    $\Omega_{\rm{B}}   \approx
0.06\,h_{50}^{-2}$  from  primordial nucleosynthesis  (e.g.,  Kolb  \&
Turner  1990;  Peebles  et al.   1991)  argues  for the  existence  of
baryonic DM.  With these two observations Sciama et al. concluded that
it  may be premature to  argue that the Davidsen  et al.   observation
conclusively rules out the decaying neutrino model.

Our goal in the present  paper is to test the  conjecture of Sciama et
al. (1993) against a set of reasonable assumptions.   We will do so by
deriving detailed  density profiles  for   the neutrinos using  N-body
experiments,   which  have   initial   density  fluctuation amplitudes
consistent  with  the  recent COBE   measurement (Smoot et al.  1992).
Within this   framework, this  will    enable   us to  make   definite
conclusions as to whether the neutrino DM  is capable of providing the
principal component to the DM distribution in the core of the cluster,
and to assess the need for baryonic DM in the cluster core as required
by Sciama et al. to evade the Davidsen et al. null result.

In this paper we use $H_0=50\,h_{50}$ km s$^{-1}$ Mpc$^{-1}$.

\section{Dark Matter in Cluster Cores}

Recent X-ray  observations of  clusters of galaxies  imply that the DM
distribution  tends to be  more centrally  condensed  than either  the
galaxy  distribution or the  distribution   of the hot   intra-cluster
plasma (see, e.g., Sarazin 1992).  In this section we  wish to discuss
the arguments for neutrino DM in the central cores of galaxy clusters,
and in detail consider a recent model for the mass distribution in the
cluster A665 which argues in favor of a  baryonic  component to the DM
distribution in the core of the cluster A665.

Given the collisionless  nature of the  neutrinos it is  often thought
that the  neutrinos  are unable to  fall deep into  the  cluster core.
Doroshkevich et al.  (1981) showed that when $k_{\rm B} T_\nu << m_\nu
c^2$ the characteristic velocity of the  neutrinos is  roughly $6.6 \,
(m_\nu/30\,{\rm eV})^{-1}\,(1 + z)$   km sec$^{-1}$,  which implies  a
characteristic  temperature  of   $4.6  \times 10^{-5} (m_\nu/30\,{\rm
eV})\, (1 +  z)$  K for  a neutrino mass of  $30$ eV. This compared to
typical dispersion velocities in clusters of order several thousand km
sec$^{-1}$. Thus, the characteristic temperatures  of the neutrinos is
much smaller than  the galaxies, and it appears  that there may  be no
energetic  problem with the  neutrinos  being trapped  in  the cluster
potential   well if  the  latter is varying  with  time in a  suitable
fashion (e.g., violent relaxation).

In 1979 Tremaine  \& Gunn showed   that one can   place  limits on the
neutrino mass   by phase  space  arguments.  One  might  be tempted to
believe that there might exist  phase-space limits to the  density  of
neutrinos in   the cluster core,  but   if  we use  their limit on the
neutrino mass

$$
m_\nu > 101 {\rm eV} \left( {100 {\rm km/s} \over \sigma} \right)^{1/4}
\left( { 1 {\rm kpc} \over r_{\rm c}} \right)^{1/4} g_\nu^{-1/4},
$$

\noindent where $\sigma$ is the the velocity dispersion and $r_{\rm c}$
is the core radius, and assuming typical quantities for a rich cluster
(for instance $r_{\rm c} \approx 0.25$  Mpc, and $\sigma \approx 10^3$
km/sec) we obtain a rather unrestrictive bound on the neutrino mass of
$m_\nu >  3.6 {\rm eV}  g_\nu^{-1/4}$.   Thus there  do not exist  any
phase space limitations on neutrino DM in the core of a cluster.

Important advances have recently occurred in the understanding  of the
mass  distribution in clusters  of galaxies.  In this  section we will
focus, in detail, on a recent mass model of A665.

Assuming hydrostatic equilibrium, X-ray data  are in principle able to
give the cluster's binding mass at any radius,  $M(r)$.  Given the gas
density ($\rho_{\rm g}$) and temperature ($T$) profiles the expression

$$
M_{\rm t}(r)~=~ {k_BT \, r \over G \mu m_p}~\biggl( {d{log}\, \rho_g \over
d{log}\,r} ~+~ {d{log}\, T \over d{log}\,r} \biggr)
$$

\noindent gives $M(r)$ where all the quantities have their usual
meaning  (for  a review  see   Sarazin 1986).   While  previous  X-ray
satellites   lacked sufficient resolution to give   $T(r)$, the latest
generation of detectors (from, e.g.,  the {\it Ginga}  and {\it Rosat}
satellites) provide both $\rho_{\rm g}(r)$  and $T(r)$.  Thus, one can
separately deduce  the mass  in  the hot  intracluster plasma  $M_{\rm
g}(r)$,    and  the  total  (binding)   mass,   $M_{\rm t}(r)$.    The
distribution of mass locked in galaxies is  usually estimated from the
observed number  counts distribution assuming  an average  galaxy mass
(see for instance The \& White 1986).  Recent observations of a number
of clusters have yielded  a striking and  unexpected  result:  the  DM
distribution is characterized by a smaller core  radius and  a steeper
slope than is the case for the gas and  the galaxy distribution (Eyles
et at. 1991; Briel et al. 1992; Gerbal et al.  1992;  Hughes \& Tanaka
1992).  This is exactly the opposite of what is observed  in galaxies,
where the DM is detected  to  be substantially more spatially extended
than the visible matter (see   Sciama  et  al.  1993),   and could  be
interpreted as favouring dissipational baryonic DM in clusters.

For  the cluster A665, we  now   discuss a self-consistent mass  model
derived from Hughes  \& Tanaka (1992) and based  on   a combination of
X-ray ({\it Ginga}) and optical data. The model includes:
\smallskip

\noindent
1) A galaxy component of spatial density distribution,
assuming ($M/L_V=20\,h_{50}$),

$$
{\rho_{\rm G}(r) \over \rho_{\rm crit}} ~=~
{750 \over 1 ~+ (r/530 \,{\rm kpc})^2 }\,.
$$

Note that  Hughes and Tanaka assumed  that the galaxies in the cluster
have   $M/L_V=5$.  However, the   extended halos   around spirals  and
ellipticals   certainly  warrant  a higher   value.  We assume  a more
realistic $M/L_V=20$ (e.g., Broeils 1992).   This change increases the
DM central density spike.
\smallskip

\noindent
2) A hot-gas component,

$$
{\rho_{\rm g}(r) \over \rho_{\rm crit}}  ~=~
{1500 \over 1 ~+ (r/380\, {\rm kpc})^2 }\,,
$$

\noindent and the total (binding) mass distribution is
(Hughes \& Tanaka 1992):

$$
{\rho_{\rm t}(r) \over \rho_{\rm crit}} ~=~
{19700 \over [1 ~+ (r/298\, {\rm kpc})^2]^{1.36} }\,.
$$

The  DM density profile   can be obtained    from $\rho_{\rm  DM}(r) =
\rho_{\rm t}(r) - \rho_{\rm g}(r) - \rho_{\rm G}(r)$.  The  DM density
profile is plotted in Figure 3, and  from the figure  it is clear that
there  is strong  evidence   for a  substantial  DM  component  in the
cluster. Furthermore  this DM component has a  smaller core radius and
is steeper  than  the corresponding density profile   for  the visible
components. Equivalently, one might say that, compared  to the visible
mass components, the DM profile shows a central spike.

\section{Simulations}

The PM simulations (Hockney \&  Eastwood 1980) were 128$^3$  particles
in  128$^3$ cells  with the  Hot Dark  Matter  (hereafter HDM) initial
power spectrum  (e.g.,   Bond and  Szalay   1983) with power  spectrum
$P(k)\propto  k$ at  small   $k$.    The   amplitude of the    initial
perturbations  is  chosen to give  the  nonlinear autocorrelation $\xi
(r)\propto r^{-1.8}$ at the moment we choose  for our analysis so that
it  is  in agreement with  observational  data on  galaxy correlations
(Peebles 1980).  It is an  extremely interesting coincidence that this
normalization predicted the $({\Delta T/T})_{RMS}
\approx 10^{-5}$ recently observed by COBE (Melott 1984b; Anninos
et al.   1991).   Predictions of higher   amplitudes for the microwave
background  fluctuations in  HDM   models have  been  based on  linear
theory, not full nonlinear N-body simulations.

There   are, of course,   well--known  problems  with  the  HDM model.
Galaxies  are  expected to form only  in  the filamentary or sheetlike
``pancakes" that form (Fig.   1), because compression takes place only
in  these sites.  When this  restriction is  made, the autocorrelation
amplitude of  the material  that remains is too  high to be compatible
with that of galaxies  (White et al.  1983).  Therefore we must appeal
to  some other process  than gravitational  instability in the pancake
theory to make  galaxies,  for  example some  sort  of distribution of
seeds. Hopes that radiation pressure from the decaying neutrinos might
start the (Hogan 1992) process have not worked,  basically because the
$uv$ ionizes the universe more efficiently than  it heats it (Splinter
and  Melott 1992).  Perhaps a  model in  which  galaxies are seeded by
independent   fluctuations is necessary    (Scherrer  et  al.    1989;
Villumsen et al.   1991).  At any rate, we   take the COBE  amplitude,
which produces  a  mass autocorrelation slope and amplitude compatible
with the present galaxy correlation amplitude, and beg the question of
galaxy formation. The result will be a  lower  bound to the DM density
in clusters, since additional power introduced on  small  scales would
intensify the clustering.

In order to  clearly  understand  any  impact  of  boundary conditions
(finite box size) or dynamical resolution, we simulated  three cubical
volumes, one  32 Mpc $h^{-2}$,  one 64  Mpc  $h^{-2}$ and one  128 Mpc
$h^{-2}$ on a side. Any interaction  with computational volume effects
should betray itself by giving   a systematical difference in  cluster
properties.

Since  this model  is not  hierarchical,  but has a  truncated initial
power spectrum, there  are    discontinuities  in the  description  of
objects corresponding to  voids, sheets,  filaments,  and clusters. We
were    able to   identify clusters  by  looking   for density  maxima
$\rho/\rho_{\rm c}>100$ and examining their properties. Since caustics
are not resolved, they do not complicate the process.

After locating the highest density region of each  cluster,  we binned
density   into spherical shells  surrounding  that peak  to generate a
density profile.  The  properties of such simulated  clusters  will be
compared  in the  next section  with those  of  actual  clusters  (and
notably A665).

We found impressive agreement  over nearly the  entire range of radii,
except   that smaller boxes    with their  higher   resolution  gave a
different result at small $r$ outside  the dynamic range of the larger
box.   We  found  all three curves  agreed  remarkably well  down to a
radius of about 0.3 cells, considerably better  than  we expected.  We
therefore show in Fig.  2 the averaged density profile,  for the small
box down   to $r=6.25  \times  10^{-2}$ Mpc   $h^{-2}$,   the smallest
reliable  value.  The  errors are  one  standard  deviation  from  one
cluster to another in the mass density in a shell.

In agreement  with the argument suggested  in section  II on neutrinos
being  trapped in clusters,  we  see that  the neutrinos are   able to
cluster into clumps   of rather high  central  density  with  radially
extended density  profiles. An  analytic fit  to  the  average density
profile of the simulated clusters yields:

$$ 
{ {\rho}_{\rm DM}(r)
\over \rho_{\rm crit}} = { 2200 \pm 700 \over
[ {1 +  ({r / 910\,h_{50}^{-2} {\rm kpc}}
)^2 }]^{1.17}} \eqno(1)
$$

where the  uncertainty  represents the  estimated   range  of  central
densities for our set of different cluster realization (see Fig.3).

In  Fig. 3 we  plot the   individual neutrino density distribution  of
HDM--simulated clusters (solid  lines) vs  the "observed"   DM density
distribution of  A665 (open squares). This   predicted density profile
appears to agree well with the  distribution of  mass in A665 at small
radii. At large radii there are large discrepencies, but note at these
large radii  the observations have  large errors.  This is despite the
fact that we assumed an initial amplitude for the density fluctuations
that is  consistent with   the  COBE observation.  If,    as suggested
recently, some of the observed COBE amplitude  might in part be due to
gravitational  waves (see, e.g.:  Davis  et al.   1992; Liddle \& Lyth
1992;  Lidsey \& Coles   1992;   Lucchin  et  al. 1992; Salopek  1992;
Souradeep \& Sahni 1992), then  the amplitude of  our density profiles
could  be  lowered,    making  them  somewhat   more compatible   with
observations.  In any case the possibility of {\it two}--components to
the  DM distribution  remains  open:  a   baryonic  component  with an
overdensity greater than $10^4$ dominating the innermost central $300$
kpc,  plus an  exotic   component (e.g., the   neutrinos  described by
eq.[1]) dominating elsewere,  fits  DM  density distribution  in  A665
quite well,  but note a   large  baryonic  contribution  to   the   DM
distribution  is not   necessarily required as   demonstrated by   our
simulations.  Therefore the neutrino interpretation  of the cluster DM
is still viable in the light of recent observational progresses and it
is pertinent to investigate the limits on  the timescale decay implied
by the Davidsen et al. (1991) null result.

\section{Decaying Neutrinos}

The neutrinos contributing to the  flux observed  by HUT are contained
in  the volume defined  by sliding the HUT  window along the  line  of
sight from one edge of the cluster to the other. The actual expression
of the density distribution  is required for  an  accurate estimate of
the total flux:  since two  dimensions of the  volume  considered  are
quite small (and in any case comparable with the core radius of the DM
distribution),  the  value of  the  central density  is important; and
since the  third dimension extends  to  the   cluster edge, the radial
profile of the DM plays an important role as well. In detail, in order
to compute the DM mass comprised within the cluster  volume accessible
to HUT, we  have   integrated the  DM    distributions,  corresponding
respectively to our A665 model and to our simulation,  over the volume
defined by the  rectangular  HUT  window ($A_{\rm HUT}=68  \times 457$
kpc$^2$ at the distance of A665) and by a radial edge of 3 Mpc.

Assuming first  that all the  DM resulting from our mass decomposition
is made by decaying   neutrinos, we  obtain   a mass similar  to  that
estimated by Davidsen et al. (1991). Since the flux is proportional to
the mass (see below), from our mass  model we predict  a flux which is
also similar  to (slightly lower than) that  estimated  by Davidsen et
al. (1991), which similarly implies a lower limit
on the decay time of 

$$
\tau_{23} > (6 -20)\, \left( {29 \ eV} \over m_\nu \right)
$$

\noindent (with  $\tau_{23}$ the neutrino lifetime  in units of $10^{23}$ secs),
also in accord with Davidsen et al. (1991).

On the other hand, according to  our simulations  the neutrino density
in clusters  follow the mean   relationship   given  by  eq.(1).  This
implies,  by the   same   procedure as used above,  a  total  mass  in
neutrinos (integrated over the HUT slit) about 2  times lower than the
Davidsen et al. value: $(1.3 \pm 0.4)
\times 10^{13} M_\odot$. So the predicted neutrino flux from the simulated
cluster taken  to  be representative of   A665 will be correspondingly
lower than Davidsen et al.'s expected value:

\begin{eqnarray}
F_{\nu}~ = ~ (0.039 \pm 0.013) ~
\biggl( {A \over 0.068\, {\rm kpc} \times 0.457 \, {\rm kpc}} \biggr)
\biggl( {f \over 1.5} \biggr) 
~ \biggl( {\rho_{\rm o} \over 2000 \, \rho_{\rm c}} \biggr)
\biggl( {R_{\rm core} \over 0.3} \biggr)
\biggl( {\epsilon \over 14\,{\rm eV}} \biggr)^{-1}
\tau_{23}^{-1}
\end{eqnarray}

\noindent where $A$ is the area of the detector slit $f \simeq 1.5$ accounts for
the radial variation of the density inside the HUT window, $\rho_0$ is
the central density, $R_{\rm core}$  is the DM  core  radius  in  Mpc,
$\epsilon$  is the monochromatic  energy of   the   decay photons,
$\tau_{23}$ is the neutrino lifetime in  units of $10^{23}$ s, and 
$F_{\nu}$ is in units of $ {\rm photons} \ {\rm s}^{-1} \ {\rm cm}^{-2}$.  
Scaling
the Davidsen et al. value of $\tau_{23}$ to the mass  (or flux) of our
simulated cluster  we   then get a  tighter lower  limit for the decay
timescale:

$$
\tau_{23} > (3 \pm 1) \, \left( {29 \ eV} \over m_\nu \right)\,.
$$

This limit is a factor 2  smaller than the  one originally obtained by
Davidsen et al. (1991), in that while they assume that all the central
DM is neutrino for   the purpose of computing  the  flux, we  take the
alternative view of estimating the minimal  flux that  in our scenario
neutrino must  emit from this cluster,  independently of whether  some
baryonic DM resides in their  innermost regions. We emphasize at  this
point  the  possibility that   baryonic DM,  although  quite  a  minor
component  by mass fraction cluster-wide,   might  still dominate  the
inner  $\sim$100  kpc  and show  up as   a central density  spike. Our
constraint, i.e. strictly $\tau_{23} > 2 (m_\nu/29\,  {\rm eV})^{-1}$,
can  be compatible  with part of Sciama's (1990)  range  of values for
$\tau_{23}$,   and of    course easily  the  original  (Melott  1984a)
suggestion of $10^{24}$ s.

\acknowledgements

Computations were performed at the National  Center for Supercomputing
Applications,  Urbana,  Illinois.   ALM   gratefully  acknowledges the
support of NSF  grants   AST--9021414 and   NSF   EPSCoR Grant  number
OSR-9255223 and NASA  grant NAGW--2923. RJS would  like to thank  NASA
for providing support through a NASA Graduate Student Fellowship.

\begin{figure}

\caption{A plot of the particles in a thin slice of our medium
simulation of a neutrino--dominated universe.}

\caption{The average density profile of DM in
clusters in our high resolution box (box size 32 Mpc). The horizontal lines
indicate one
standard deviation in the density.}

\caption{The DM density distribution of A665 (open squares)
and the neutrino density distribution from our high resolution simulation
(solid lines).  Observations are uncertain for  $R > 2 {\rm Mpc}$,
and the simulation can not resolve the innermost $\simeq 500$ kpc.}

\end{figure}

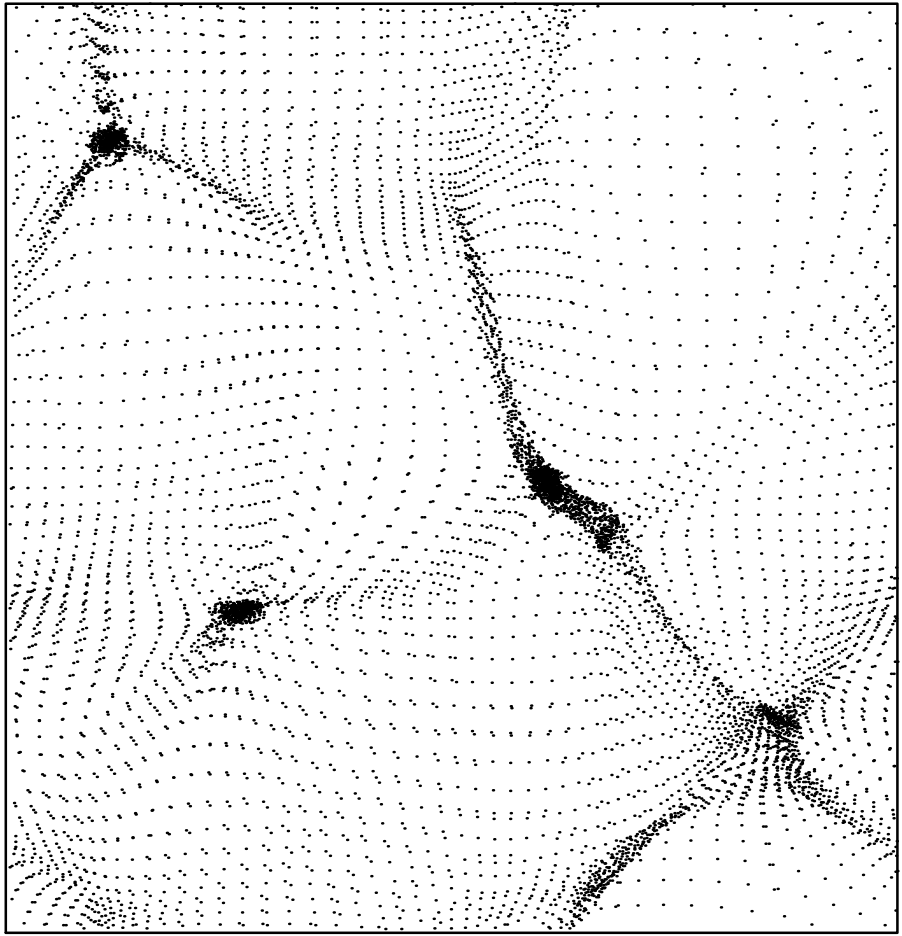
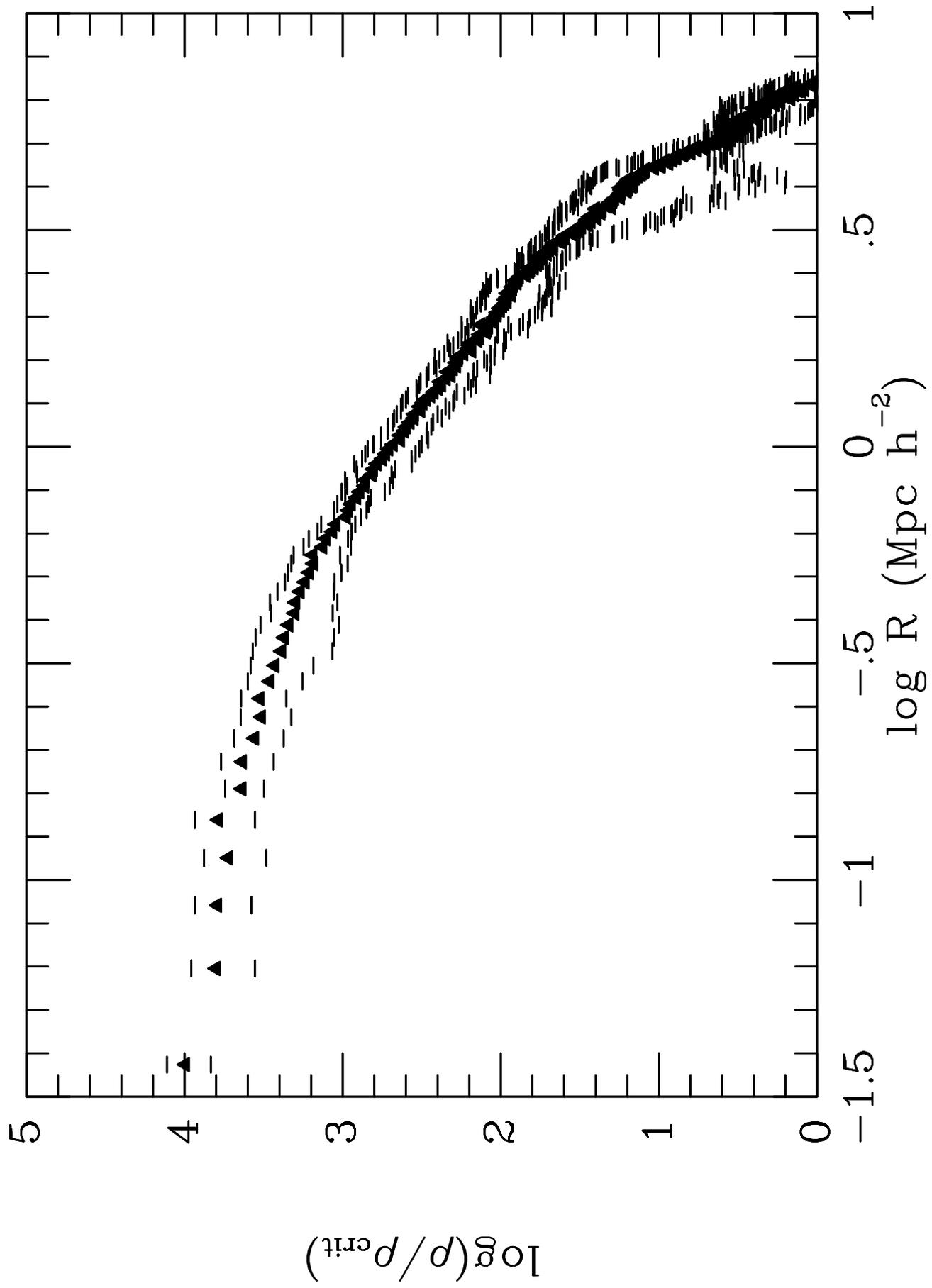
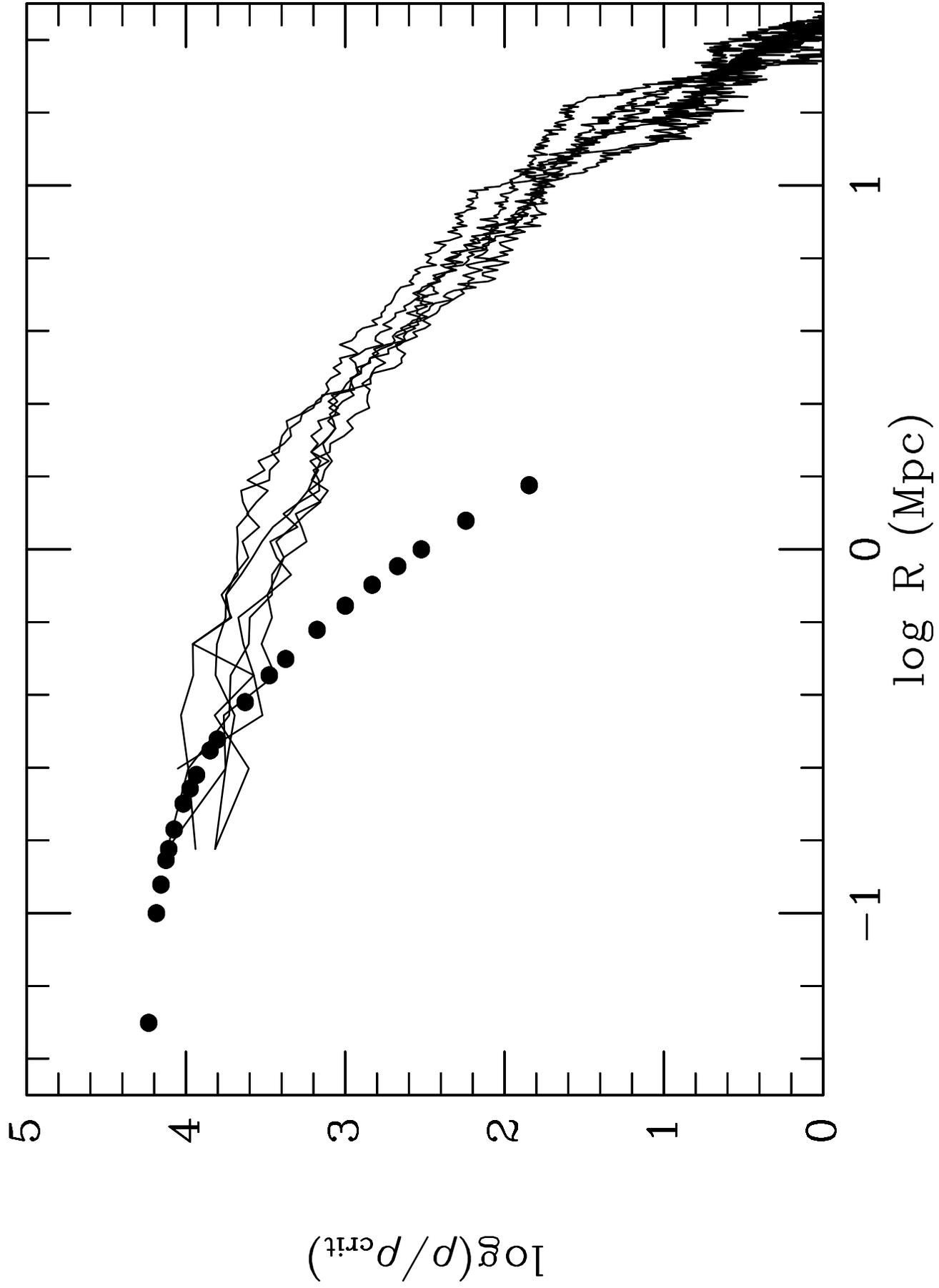
\end{document}